\documentclass[aps,pre,twocolumn,floatfix]{revtex4}
\pdfoutput=1
\usepackage{graphicx}
\usepackage{color}

\begin{document}

\title{
Probing the entanglement and locating knots in ring polymers: a comparative study of different arc closure schemes}

\author{
Luca Tubiana$^1$, Enzo Orlandini$^3$, Cristian Micheletti$^{1,2}$
}

\affiliation{
  \centerline{$^1$ International School for Advanced Studies, Via Bonomea 265,
  I-34136 Trieste,
  Italy}
  \centerline{$^2$ CNR-IOM Democritos and Italian Institute of Technology (SISSA
  unit)} \centerline{$^3$ Dipartimento di Fisica ``G. Galilei'' and Sezione INFN,
  Universit\`a di Padova, Via Marzolo 8, I-35100 Padova, Italy }}

\begin{abstract}
  The interplay between the topological and
    geometrical properties of a polymer ring can be clarified by
    establishing the entanglement trapped in any portion (arc) of the
    ring.  The task requires to close the open arcs into a ring, and
    the resulting topological state may depend on the specific closure
    scheme that is followed. To understand the impact of this
    ambiguity in contexts of practical interest, such as knot
    localization in a ring with non trivial topology, we apply various
    closure schemes to model ring polymers. The rings have the same
    length and topological state (a trefoil knot) but have different
    degree of compactness. The comparison suggests that a novel
    method, termed the minimally-interfering closure, can be
    profitably used to characterize the arc entanglement in a robust
    and computationally-efficient way. This closure method is finally
    applied to the knot localization problem which is tackled using
    two different localization schemes based on top-down or bottom-up
    searches. 
\end{abstract}

\maketitle

\section{Introduction}

It is known that the global topological state of a ring polymer
affects its salient physical properties such as its size
~\cite{desCloizeux:1981:J-Phys-Lett,Moore:2004:PNAS} sedimentation
velocity, gel-electrophoretic
mobility~\cite{Stasiak:1996:Nature:8906784,
Weber_et_al_2006_Biophys_J,Orlandini_et_al_2010_PRE}, resistance to
mechanical stretching~\cite{Saitta_et_al:1999:Nature} or the behaviour
under spatial confinement
~\cite{Marenduzzo:2009:Proc-Natl-Acad-Sci-U-S-A:20018693}.

While a comprehensive understanding of this phenomenon is still
lacking, it is often explicitly or implicitly acknowledged that
topology-dependent physical properties arise because of a
sophisticated interplay of polymer geometry and topology. In other
words, the global topological state affects the average
geometrical properties of the polymer, which in turn directly
impact various physical properties such as those mentioned above.

A vivid illustration of this relationship is offered by the mechanical
resistance of a knotted polymer that is pulled at both ends. The
breaking force depends on the topological state of the
polymer. Indeed, the rupture point is invariably in
correspondence of the knot~\cite{Saitta_et_al:1999:Nature} which is
progressively tightened by the pulling action (as all fishermen know
for the case of a knotted fishing line).

The above example highlights a very important player
  in the relationship between the topological, geometrical and
  physical properties of a ring polymer (or a polymer with constrained
  ends), namely the degree of localization of the
  topologically-entangled
  region~\cite{Orlandini_et_al_2009:Phys-Bio}. For example, recent
  simulations have shown that the delocalization of ``knots'' in a
  linear DNA filament is very important to allow its tranlocation
  through a pore (as in viral DNA ejected from the viral capsid)
  avoiding plug-like
  obstructions~\cite{Marenduzzo:2009:Proc-Natl-Acad-Sci-U-S-A:20018693}.

Locating the knotted portion of the polymer is
  straightforward when the knot is tight, but is otherwise highly
  challenging.  Generally speaking, to accomplish this task one needs
to establish the degree/type of entanglement ``trapped'' in any
portion, or arc, of the ring polymer and then select the shortest
arc(s) whose trapped entanglement matches the global topology of the
ring. The entanglement trapped in a given arc is identified by
establishing the topology state of an auxiliary ring obtained by
suitably joining the two arc ends.

The difficulty of this scheme lies in the fact that
  several viable arc closure (end-joining) schemes can be used and
  they can result in different knots being measured on the same arc. This is especially
  the case for rings under geometrical
  confinement~\cite{Millett_2005_Macromol,Orlandini&Whittington:2007:Rev-Mod_Phys,Marcone:2007:PRE}.

While the above-mentioned ambiguity is ultimately
  unavoidable, it is important to ascertain how severe it is in
  contexts of practical interest. This question, motivates the present
  study where we use and compare several closure schemes to
  characterize the entanglement trapped in portions of three model
  ring polymers with the same length and knotted state (trefoil
  knot) but with different degree of compactness and hence of
  geometrical complexity.
  \cite{MichelsWiegel1986,Micheletti:2006:J-Chem-Phys:16483240,Micheletti:2008:Biophys-J:18621819,Baiesi_et_al:2009:JCP,Panagiotou_et_al:2010:J-Phys-A}.

Three different closure schemes are considered: the \emph{direct
  bridging closure}, the \emph{stochastic closure} and the novel
\emph{minimally-interfering closure}.  The first two have been
introduced previously~\cite{Marcone:2005:J-Phys-A,Millett_2005_Macromol},
while the third is presented and applied here for the first time.

From the comparative investigation we ascertain that for the ring with
the least degree of compactness (spatially unconstrained), the various
closure schemes yield consistent results for the entanglement trapped
in the various arcs. For the higher level of compactness noticeable
differences emerge between the direct bridging method and the other
two schemes. Notably, despite their different
formulation, the stochastic closure and the computationally faster
minimally-interfering closure appear to be well consistent for all the
considered levels of ring compactification. This is an important
result as it gives an {\em a posteriori} indication of an overall
consensus of unrelated methods about the topological state of various
portions of rings with different geometrical complexity.

The implications of the findings for the problem of
  knot localization are finally discussed. Specifically, we consider and
  compare two alternative knot localization procedures previously
  discussed in the
  literature and corresponding to top-down and bottom-up searches~\cite{Katrich:2000:PRE,Marcone:2005:J-Phys-A,Mansfield2010}. We show
  that the geometrical complexity of the ring resulting from
  increasing confinement of the rings correlates
  with appreciable differences in the regions that are identified as
  accommodating the knot.

\section{Methods}

\subsection{Polymer model}

The degree of entanglement is measured for the simplest model of ring
polymers, that is freely-jointed rings (FJR).  These rings are
fully-flexible equilateral polygons and no excluded volume interaction
is introduced between the ring edges or vertices.

It is known that the global topological complexity of the
rings is strongly influenced by the level of imposed
spatial confinement. Typically, a higher degree of ring compactification
leads to more complex knots. This aspect was initially investigated by
Michaels and Wiegel~\cite{MichelsWiegel1986} and more recently by
other
studies~\cite{Arsuaga:2005:Proc-Natl-Acad-Sci-U-S-A:15958528,Micheletti:2006:J-Chem-Phys:16483240,Micheletti:2008:Biophys-J:18621819,Marenduzzo:2009:Proc-Natl-Acad-Sci-U-S-A:20018693}
in biologically-motivated contexts, see refs. \cite{Review2010} and
\cite{review2011} for two recent reviews.

It is therefore envisaged that, by focusing on conformations having a
specific topological state (such as trefoil knots) and different
degree of compactness, one might observe a very different level of
geometrical complexity, i.e. local entanglement, associated to the
same knot type.

We have consequently mapped in detail the topological entanglement for
all subportions of three equilateral rings of $N=100$ edges of unit
length. The ring configurations are picked randomly from a pool of
Monte-Carlo equilibrated structures subject to three different
isotropic confining pressures.  Specifically, one configuration was
picked from the unconstrained ensemble (zero confining pressure),
which is largely dominated by unknotted rings.  The radius of the
smallest sphere that is centred on the ring centre of mass and that
encloses all ring vertices is $R_c=4.8$.

The second configuration has enclosing hull radius equal to $R_c=
4.1$. This hull radius is close to the value of $R_c$ for which the
probability of observing a trefoil in rings with $N=100$ edges is
maximum, see ref. \cite{Micheletti:2006:J-Chem-Phys:16483240}.  The
third configuration has hull radius equal to $R_c=2.5$ and was picked
at values of the confining pressures so high that the knot spectrum
was dominated knots with topology more complex than the trefoil one.

\subsection{Closure schemes}
\label{closure_schemes}

\begin{figure*}
\includegraphics[width=\textwidth]{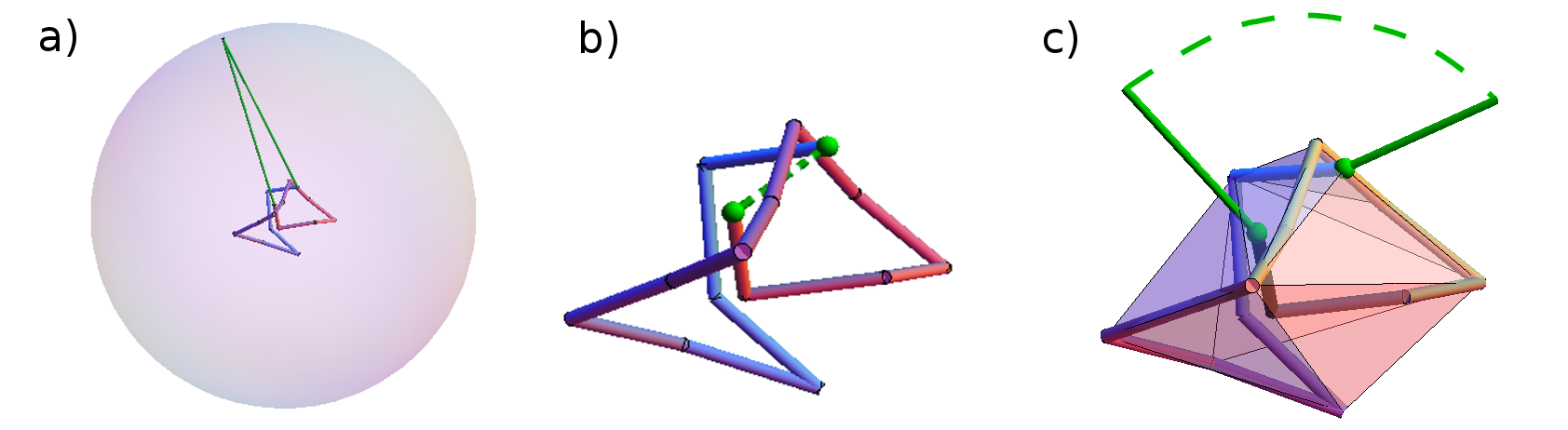}
\caption{ Sketched examples of the closure schemes used in this work: 
(a) Stochastic closure at infinity, (b) direct bridging,
(c) minimally interfering closure.} 
\label{fig:chiusure}
\end{figure*}

As anticipated in the introduction, one needs to introduce a
well-defined procedure to close, or circularize, various subportions
of the ring under consideration so as to properly establish their
topological state.

Several viable closure schemes are considered, including some that
have been introduced and applied before.  

Before describing the various closure schemes we clarify the notation
that will be used from now on. For a given ring,
$\Gamma=\{\vec{r}_1,\vec{r}_2,\ldots,\vec{r}_N,\vec{r}_{N+1}=\vec{r}_1\}$,
we denote by $\Gamma_{ij}=\{\vec{r}_i,\ldots,\vec{r}_j\}$ the arc
comprising all edges from vertex $i$ to vertex $j$
(including the endpoints $i$ and $j$). The
orientation of $\Gamma_{ij}$ follows the one given by the increasing
numbering of the nodes on the full ring.
 
The three following schemes are used to close a given arc $\Gamma_{ij}$:
\begin{small}
\begin{enumerate}
\item {\bf Direct bridging.} The two arc ends, $i$ and $j$ are
directly joined by a straight segment 
(Fig.~\ref{fig:chiusure} (b))~\cite{JansevanRensburg:1992:J-Phys-A}.
\item {\bf Stochastic closure at infinity.} This scheme, introduced in
  ref. \cite{Millett_2005_Macromol} consists in closing
  $\Gamma_{ij}$ by connecting its ends to a point picked randomly on a
  sphere centred at the centre of mass of the arc and with radius much
  larger than the radius of gyration of $\Gamma_{ij}$, see
  Fig.~\ref{fig:chiusure} (a).  The random closure procedure is
  repeated a large number of times (1000 in this study), and the
  topological state with the largest statistical weight is identified.
  If this weight exceeds a preassigned threshold, $q$, then the
  dominant topological state is taken as the topological state of the
  arc $\Gamma_{ij}$. Otherwise, the topology of $\Gamma_{ij}$ is
  considered ambiguous and is left undetermined. In this study we
  consider two different threshold values: $q=50\%$ and $q=90\%$.

\item{\bf minimally-interfering closure} The amount of
    entanglement introduced by closing the arc $\Gamma_{ij}$ is
    intuitively expected to grow with the distance spanned by the
    added closing segments inside the convex hull of $\Gamma_{ij}$. In
    order to minimize such ``interference'' we consider two
    alternative closures. In the first closure, the two endpoints of
  $\Gamma_{ij}$ are prolonged through their nearest points on the
  convex hull (computed with the QuickHull algorithm
    \cite{Barber_et_Al_quickhull_1996}) and connected with an arc at
  infinity. The sum of the distance of each of the two endpoints from
  the closest point on the convex hull, $d^{out}_{ij}$ is taken as the
  measure of the associated. This quantity is compared to the
  geometrical distance of the two points, $d^{in}_{ij}$, which is a
  measure of the interference associated to the direct
  bridging closure. The minimally-interfering closure
  is picked by comparing $d^{out}_{ij}$ and $d^{in}_{ij}$. If
  $d^{in}_{ij}>d^{out}_{ij}$ then the enpoints are joined using their
  prolongations to infinity (Fig.~\ref{fig:chiusure}(c)), otherwise
  they are directly bridged (Fig.~\ref{fig:chiusure}(b)).
\end{enumerate}
\end{small}

A pleasing feature of the
minimally-interfering closure is that when the boundary of the convex hull 
of an arc contains the arc termini, then it is straigthforward to check
whether the hull and the arc realize a ball-pair. For this check one should 
ascertain that the convex hull is not pierced by the edges of the rest of
the ring.  In addition, the
closure appears appropriate for deeply buried termini too which, as
intuitively expected, will be directly joined.

\subsection{Knot localization schemes}
\label{knot_localization_schemes}

A few different procedures have been proposed so far
  to localize the shortest knotted portion of a ring with non-trivial
  topology~\cite{Katrich:2000:PRE,Marcone:2005:J-Phys-A,Mansfield2010}. They
  can be divided into two main categories depending on the strategy
  used to search for the shortest knotted portion of a ring.  

  To illustrate the two methods, let us consider a
  trefoil knot in a ring of $N$ vertices.

The first procedure involves a bottom-up search for the knot.
  The purpose is to identify the shortest portion of the ring that has the
  same topology of the whole ring. One
  starts by considering all portions of the ring (arcs) with a small
  contour length, $l$, (small means no larger than the length
  required to tie a trefoil knot). If, after closure, none of the arcs
  is found to be trefoil knotted, then $l$ is increased by one and
  the search for a trefoil-knotted arc starts again. Clearly the
  search stops when one arc of the current contour length, $l$, is
  found to be trefoil knotted. We shall refer to this arc as the 
  \emph{shortest knotted arc}. It is important to stress that the returned shortest
  knotted arc  may correspond to an \emph{ephemeral knot}. These 
  are arcs with non-trivial topology  that are contained in longer arcs with a different
  topology (which, in turn, can be contained inside arcs with different topology etc)~\cite{Millett:2010:JKTR}.

To avoid detecting ephemeral knots one can resort to a
  second method, that involves a top-down search. In this case one
  looks for the shortest knotted portion of the ring that (i) cannot
  be further shortened without losing the knot and (ii) can be
  extended continuosly to encompass the whole ring.  To do so, one
  begins by setting $l=1$ and considers all arcs of length $N-l$ and
  discards those that are not trefoil-knotted. Then $l$ is increased
  by one unit and, inside the survived arcs, one looks for
  trefoil-knotted arcs of length $N-l$. Those that are not
  trefoil-knotted are discarded and the procedure is repeated
  until at a certain value of $l=\bar{l}$ no trefoil-knotted arc is
  found.  The trefoil-knotted arc (or arcs in case of degeneracy) that
  survided at the previous iteration step (the one(s) with length
  $N-(\bar{l}+1) $) provides the desired ring portion accommodating the knot.
  We shall refer to such arc(s) as the shortest {\em
    continuously-knotted} portion of the ring, or shortest C-knotted
  portion for brevity.

\begin{figure*}
\begin{center}
\includegraphics[width=0.8\textwidth]{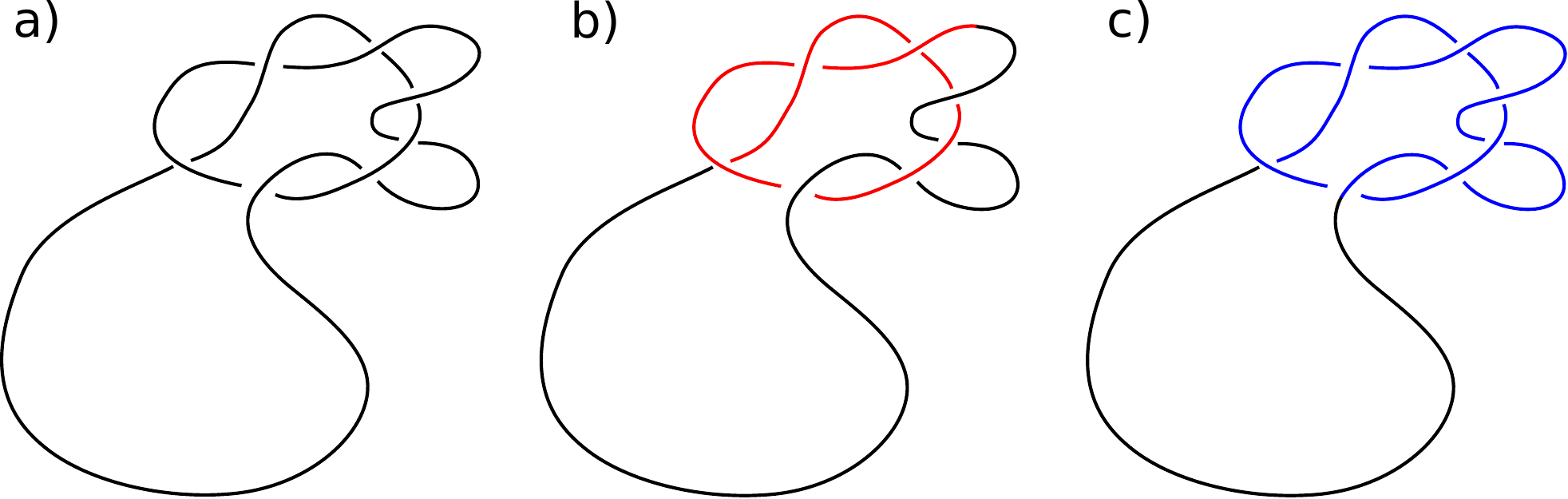}
\end{center}
\caption{ An example of how a bottom-up
    (b) and a top-down search (c) on the trefoil knotted arc shown in
    (a) give rise respectively to the shortest knotted arc (red curve
    in (b)) and to and shortest C-knotted arc (blue curve in (c)).
}\label{fig:knot_loc_diff}
\end{figure*}

  The knotted portions identified by the two procedures
  are not necessarily the same, as illustrated in
  Fig. \ref{fig:knot_loc_diff}.  
  From the two definitions it follows
  that the length of the shortest C-knotted arc cannot be smaller
  than that of the shortest knotted arc.

Both methods are applied and compared in this study,
  but with one important addition with respect to the procedure
  described above. The modification follows the observation that a
  satisfactory location of the trefoil knot in a specific arc,
  $\Gamma_{ij}$, of the ring should be accompanied by the condition
  that the complementary arc, $\Gamma_{ji}$, is unknotted. Therefore
  the test for ``trefoil-knottedness'' in the previous schemes,
  consists in the stringent requirement that upon closure
  $\Gamma_{ij}$ is trefoil knotted {\em and} its complementary arc,
  $\Gamma_{ji}$ is unknotted.

\section{Results}

\subsection{Knot matrices}

\begin{figure*}
\begin{center}
  \includegraphics[width=1.0\textwidth]{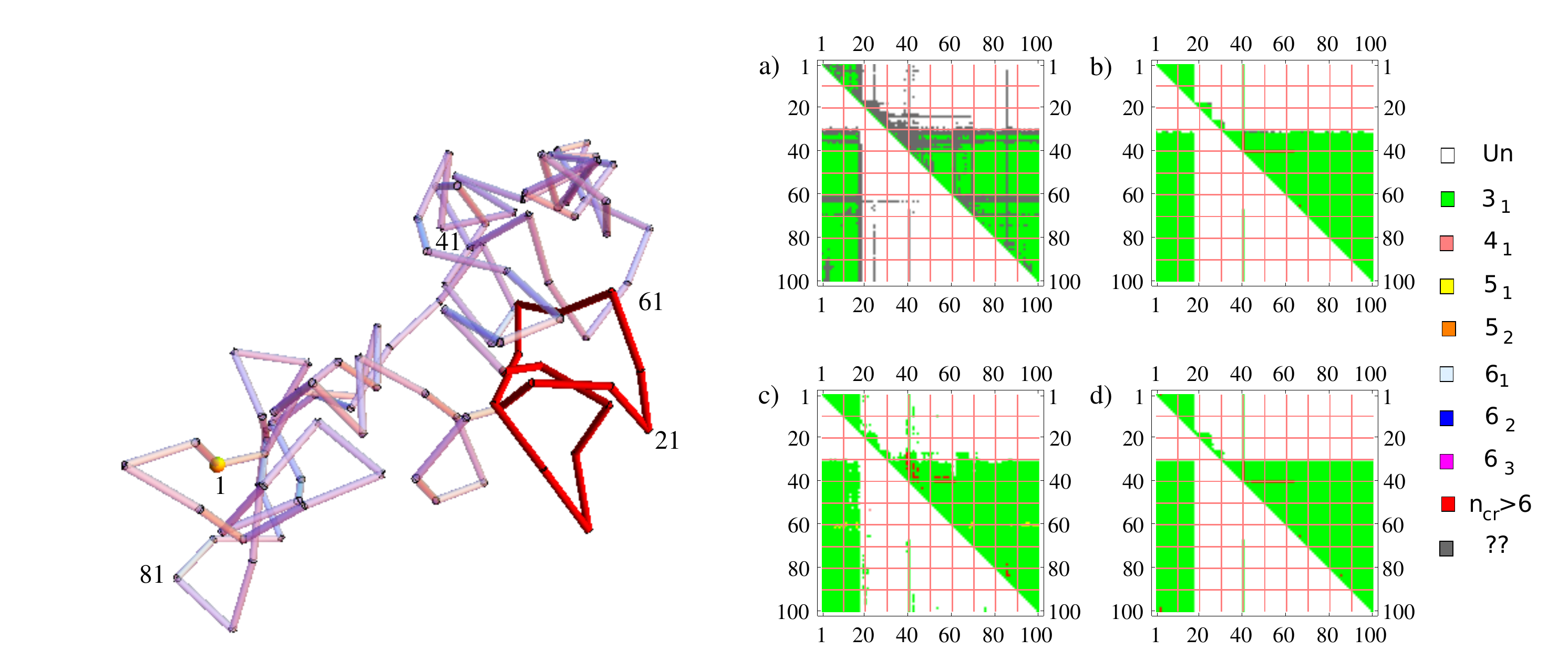}
\end{center}
\caption{ Leftmost panel: spatially-unconstrained
  trefoil-knotted ring of $N=100$ edges. The indices (numbering) of
  subset of vertices are shown explicitly; the first vertex is
  highlighted with a yellow sphere.  Other panels: knot matrices of
  the ring obtained by using $4$ different closure schemes: stochastic
  closure at infinity with threshold (a) $q=90$\% and (b) $q=50$\%,
  (c) direct bridging and (d) minimally-interfering closure. Different
  topologies are color-coded according to the legend on the
  right. The knotted portion dentified by using the
    minimally-interfering closure is highlighted in red. In this case the shortest knotted and C-knotted arcs
    coincide.}\label{fig:swollen}
\end{figure*}

\begin{figure*}
\begin{center}
  \includegraphics[width=1.0\textwidth]{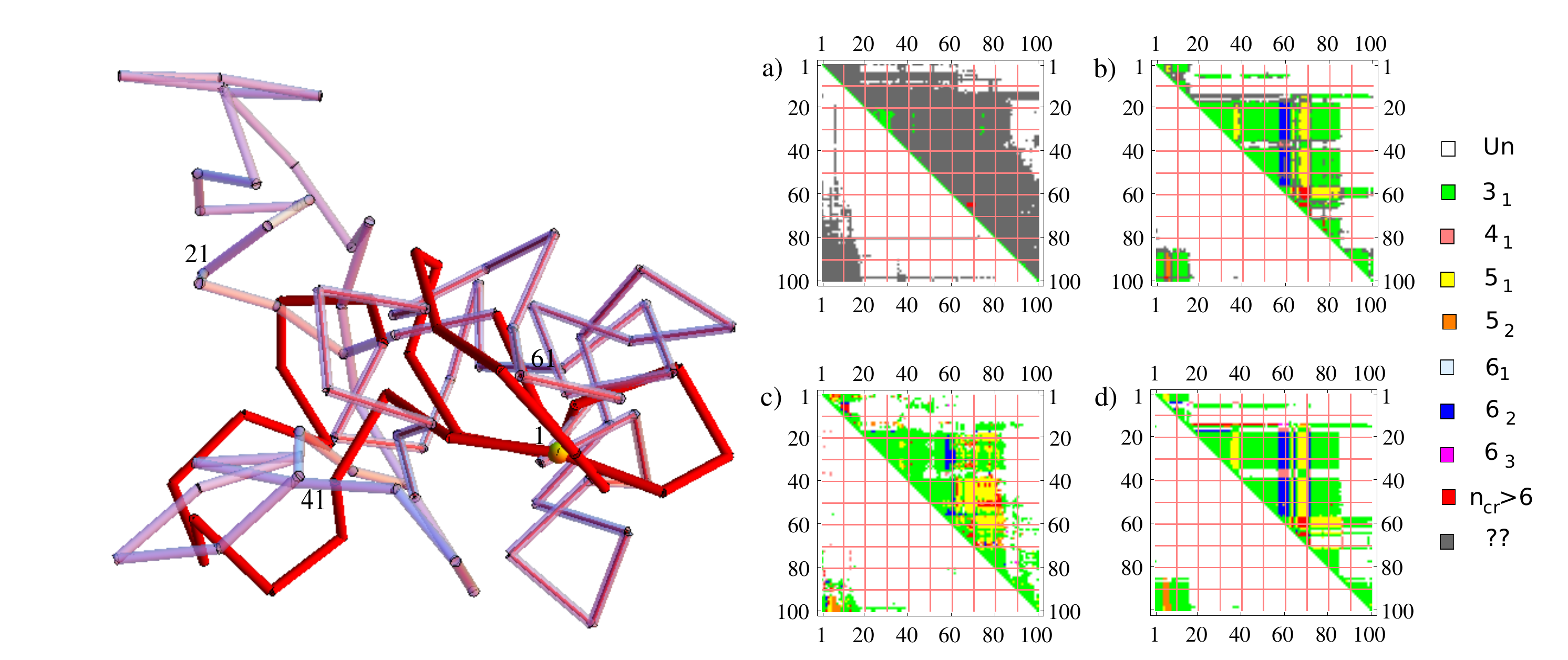}
\end{center}
\caption{ Mildly-confined trefoil-knotted ring of $N=100$ edges
(leftmost panel ) and associated  knot matrices 
  displayed and coloured as in Fig.  \ref{fig:swollen}. The shortest
  knotted arc  computed with the minimally-interfering
  closure is highlighted in red. The shortest C-knotted arc
  computed with the same closure is shown with red interior.  
  Note that the shortest continuously-knotted arc,
  $\Gamma_{51,16}$ contains the shortest knotted arc, $\Gamma_{87,15}$}\label{fig:max_prob}
\end{figure*}

\begin{figure*}
\begin{center}
  \includegraphics[width=\textwidth]{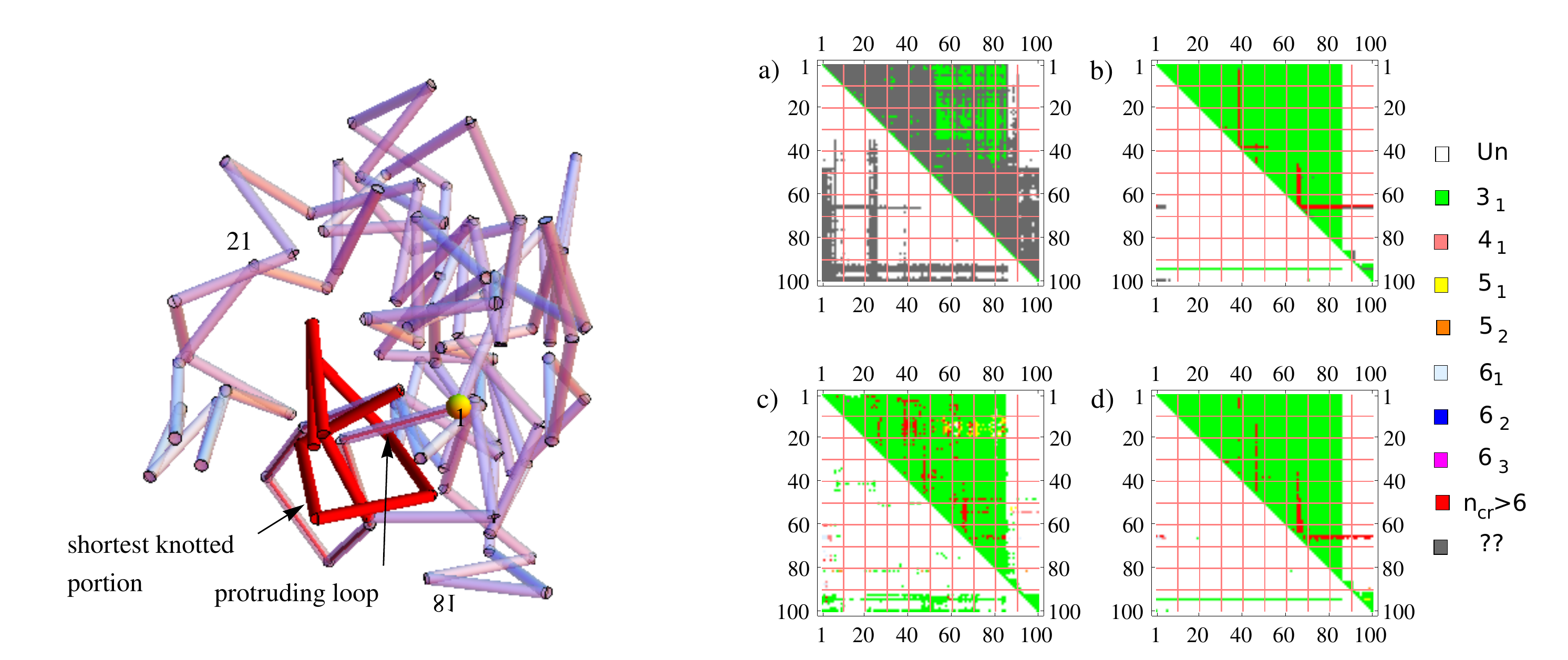}
\end{center}
\caption{ Strongly confined trefoil ring of $N=100$
  edges ( leftmost panel ) and associated knot matrices displayed and
  coloured as in Fig.  \ref{fig:swollen}. The
    shortest knotted arc computed by using the minimally-interfering
    closure (highlighted in red) is an ephemeral knot with a
    protruding loop (shown with red interior) which is contained inside a
    larger trefoil-knotted arc. The union of the shortest knotted arc and the
    protruding loop coincides with the shortest continuously-knotted
    arc.}\label{fig:min_prob}
\end{figure*}

The closure schemes described in section~\ref{closure_schemes} were
applied to all arcs of the three rings of $N=100$
edges shown in Fig. \ref{fig:swollen}, \ref{fig:max_prob} and
\ref{fig:min_prob}. 

For each ring we considered all possible $N(N-1)$
oriented arcs $\Gamma_{ij}$ with $i\ne j$.  After circularization, the
topological state of each arcs was established by using the KNOTFIND
routine implemented in the KNOTSCAPE package~\cite{Knotscape}. The
KNOTFIND routine efficiently simplifies the diagrammatic
representation of a knot and compares it against a look-up table of
diagrams of prime knots with less than 17 minimal crossings. When a
positive match is found, the topological state of the ring is
unambiguously established. If no match is found (due to genuine
excessive complexity of the knot or to insufficient classification)
the topological state is regarded as undetermined.

\begin{figure*}
\begin{center}
\includegraphics[width=0.7\textwidth]{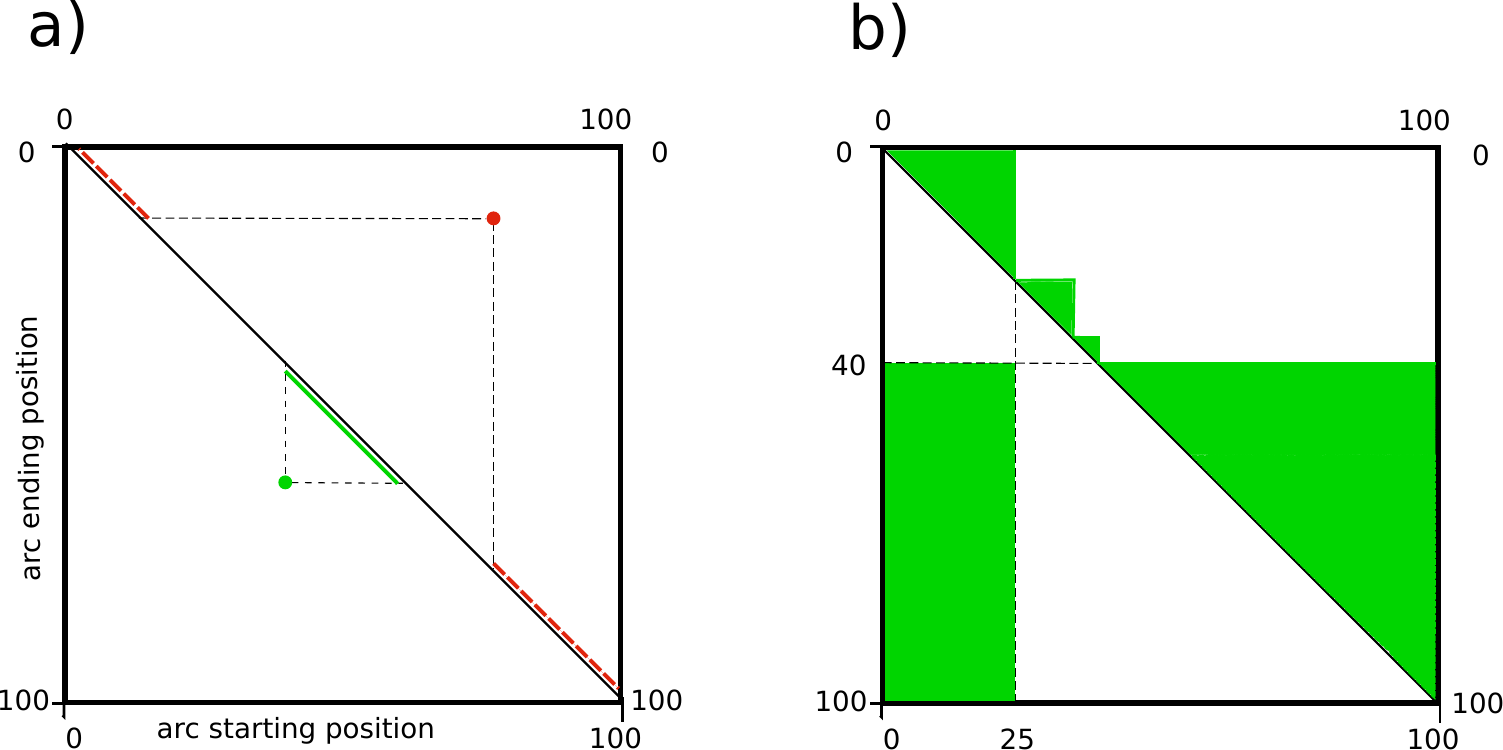}
\caption{ A ring of $N$ edges is associated to $N\times
  N$ knot matrix. Unlike the case of open chains
  ~\cite{King_et_al_2007_J_Mol_Biol} the matrix is periodic. 
  To each entry, $i,j$ of the matrix is associated an
    oriented arc $\Gamma_{i,j}$ of the ring going from $\vec{r}_i$ to
    $\vec{r}_j$.  This is illustrated in panel (a), where the arcs
    associated to the two different marked entries are highlighted on
    the matrix diagonal. The topological state of an arc is encoded by
    the color of the corresponding matrix entry. In the example shown
    in panel b white is used for the unknot and green for the trefoil
    knot. The shortest knotted arc of a knotted ring
  is the shortest subarc having the same topology of the ring
  while its complementary arc stays unknotted. For
    this ring it corresponds to the arc $\Gamma_{25,40}$ which
    coincides with the shortest continuously-knotted arc too (see
    Methods for the definition).}
\label{fig:mat_expl}
\end{center}
\end{figure*}

The topological states of all arcs $\Gamma_{ij}$ are conveniently
reported as the $(i,j)$ element of an $N\times N$ {\em ``knot
matrix''}. The knot matrix entries, which take on discrete values
reflecting the variety of knots trapped in the arcs, are conveniently
conveyed in color-coded graphical representations, see
Fig. \ref{fig:mat_expl}.  The graphical
representation adopted here follows the indexing convention first
introduced by Yeates and coworkers to highlight the presence of
slipknots in naturally-occurring
proteins~\cite{King_et_al_2007_J_Mol_Biol}. By convention the diagonal
of the matrix is taken to correspond to the whole ring.

As illustrated in Fig.\ref{fig:mat_expl} by analysing the matrix it is
possible to recover a wealth of information about the interplay of the
geometrical and topological entanglement of the ring. In particular it
is possible to identify the shortest knotted arc and the
shortest C-knotted arc as well as identifying ephemeral
  knots~\cite{Millett:2010:JKTR}.

Knot matrices have been computed for all the closure schemes presented
in (\ref{closure_schemes}). Their visual inspection and interpretation
according to Fig.~\ref{fig:mat_expl} readily conveys the salient
differences across the closure methods in establishing the degree of
entanglement of various ring subportions.

\subsection{Unconstrained case}

We start by discussing the knot matrices for the unconstrained ring in
Fig.~\ref{fig:swollen}. From an overall visual inspection, the various
knot matrices appear largely consistent and the topologies of most
arcs correspond to either unknots or trefoil knots.

Yet, as it is visible in panels (b), (c) and (d), a limited occurrence of knots
with more than $6$ minimal crossings is found for arcs of various
lengths that either start or end at vertex number $41$. For example
the arc $\Gamma_{41,39}$ is seen as a knot with more than 6 minimal crossings by all the closing
schemes. These instances are manifestly ephemeral knots because their
topological state differs from the global one of the ring,
which is the trefoil.

Note that for this ring, all schemes are 
  consistent. This fact  is compatible with the
finding of ref. \cite{Millett_2005_Macromol} that, for unconstrained
rings, the dominant knot type found with the stochastic closures with
threshold $q=50\%$ is usually the same one obtained with the direct
closure scheme.

Regarding the robustness of the closure scheme in terms of the
threshold, $q$, we report that for $q=90$\% about $15\%$ of the
entries are marked as undetermined knots (grey color). These
undetermined arcs represent cases where the details of the closure
scheme can likely yield different results. It is interesting to
observe that across panels (b), (c) and (d) most arcs whose topology
is not the trefoil or the unknot, correspond to undetermined entries
in panel (a).

It should also be noted that the direct bridging
closure scheme introduces ``jagged'' boundaries separating the trefoil
and unknotted regions. Sharper boundaries are instead found for the
stochastic closure ($q=50$\%) and the minimally-interfering
closure.

\subsection{Spatially-confined cases}

The analysis presented above was repeated for the more compact
configuration depicted in Fig.~\ref{fig:max_prob}.

The increased level of geometrical complexity compared to the
unconstrained case is conveyed by the fact that a much larger fraction
of the matrix entries ($\sim 45\%$) have an undetermined topological state
according to the stochastic closure scheme with the stringent $q=90$\%
threshold. This is because the geometrical complexity
characterizing more compact structures prevents the occurrence of a
single highly-dominant knot type.

A related aspect is that the knot matrix obtained with
  the direct bridging closure, see panel (c), is considerably noisier
that the knot matrices obtained with the tolerant ($q=50$\%)
stochastic closure and the minimally-interfering one, see panels (b)
and (d).

Notably, the visual inspection of panels (b) and (d)
indicates that the responses of these two methods remain highly
compatible notwithstanding the increased geometrical complexity.

All the above considerations hold also for the ring with highest level
of compactification shown in Fig.~\ref{fig:min_prob}. 

In summary, at all the three levels of compactness a
  high consistency is found between the tolerant stochastic closure
  and the minimally-interfering one. Given the different spirit of
  these two methods this accord is both pleasing and important because
  it provides {\em a posteriori} confidence that a consensus
  indication of the topological state of various arcs of a ring can be
  achieved with these two different closure methods.

  It is important to point out that, despite returning consistent
  results, these two methods are very different in terms of the
  computational espenditure because the stochastic closure scheme is
  based on a collection of several (in our case 1000) random closures
  per arc whereas only two closures per arc are
  involved in the minimally-interfering scheme. The latter scheme
  appears therefore to be preferable when one seeks
  to establish the local level of entanglement over a large ensemble 
  of rings or open chains (as for large-scale surveys for detecting and
 locating knots in all available structures of globular
 proteins~\cite{King_et_al_2007_J_Mol_Biol,Plos2010}.) 
    The former, however, has
  the advantage of providing a quantitative control of the statistical
  weight (and hence the robustness) associated to the dominant knot
  type for every arc.

\subsection{Locating the knot}

To locate the trefoil knot within each of the three rings
  under analysis we processed the associated knot matrices obtained by
  applying the minimally-interfering closure. We use both the top-down and
  the bottom-up approaches to locate the knot. The results are described
  herafter and are practically identical to
  those obtained by using as input the knot matrices obtained with the tolerant ($q$=50\%) stochastic
  closure.

For the unconstrained knot, the two search methods
  identify the same arc, see Fig.~\ref{fig:min_prob}, as the region
  that accommodates the knot.

This is not the case for the two more compact rings.
  In particular, for the ring shown in Fig. \ref{fig:max_prob} the shortest
  knotted arc corresponds to $\Gamma_{87,15}$ (highligthed in red in
  the figure) while the shortest C-knotted arc corresponds to the much
  longer arc $\Gamma_{51,16}$ shown with red interior.  

Finally, for the most compact ring, the shortest
  knotted arc is found to be $\Gamma_{86,95}$ while
    the shortest C-knotted arc is found to be $\Gamma_{86,1}$, see
  Fig.~\ref{fig:min_prob}. In this case the comparison between the two
  knot localization methods reveals a notable hierarchy of ephemeral
  knots. In fact, while arc $\Gamma_{86,95}$ is trefoil knotted, the longer arcs
  from $\Gamma_{86,96}$ up to $\Gamma_{86,99}$ are uknotted and still
  longer arcs, such as $\Gamma_{86,1}$, are trefoil-knotted again.

It therefore appears that the increased geometrical
  complexity of the rings resulting from the isotropic spatial
  confinement produces a non-trivial interplay of geometry and
  topology, which manifests in the sensitive dependence of the knot
  location on the search strategy that is used. A systematic study
  of the broader implications of this finding is currently
  underway.

As a final remark we point out that when dealing with
  large datasets of rings or suitably closed polymer chains, as in
  large-scale surveys of knotted proteins\cite{Plos2010,Virnaupdb}, the
  calculation of the knot detection and knot localization can be
  speeded up by an initial simplification of the ring geometry. In
  principle, such changes could affect the outcome of the knot
  localization procedure. While a systematic study of this effect is
  beyond the scope of the present investigation, we report in the
  Appendix a limited discussion of the matter.

\section{Conclusions}

The main aim of this work was to investigate the
  efficiency and reliability of different closure schemes in assigning
  a topological state to a given subportion of a ring and to
  characterize the topologically-entangled region of knotted rings
  when they are subjected to different levels of compactification.
  The detailed analysis of the local entanglement of the ring
  described in terms of knot matrices shows
  that two independent closure schemes yield robust and consistent
  results at all level of ring compactification: the stochastic
  closure and the minimally-interfering closure.

The two methods, while providing consistent results, have different
advantages. The stochastic closure scheme, in fact, provides a
statistical confidence level on the robustness of the dominant
topological entanglement associated to each considered arc. This
valuable information comes at the cost of performing a large number of
statistically independent closures on the arc of interest. The
statistical robustness information is not directly available in the
minimally-interfering scheme as this requires only
two closures. But for this very reason the minimally-interfering
method is much faster than the stochastic and yet typically returns
the same dominant knot type. Considerations of these aspects can guide
the choice of which of the two methods should be adopted in a specific
study.

The interplay of geometry and topology in rings with
  different degree of compactness was finally examined by comparing
  two different methods for knot localization (involving a bottom-up
  and top-down search, respectively). The analysis indicates that for
  spatially unconstrained rings the location of the knot can be
  performed in a consistent manner by the two methods.  Appreciable
  differences between the methods emerge for the more compact
  configurations, signalling a non-trivial increase of the geometrical
  complexity of confined polymer rings.

\section*{Acknowledgements}
We would like to thank Ken Millett for several discussions
and the organizers of the 2010 Kyoto conference on {\em Statistical
  physics and topology of polymers}

\appendix
\section{Effect of  ring simplification}
\label{chain_simp}

\begin{figure*}[ht]
\begin{center}
\includegraphics[width=\textwidth]{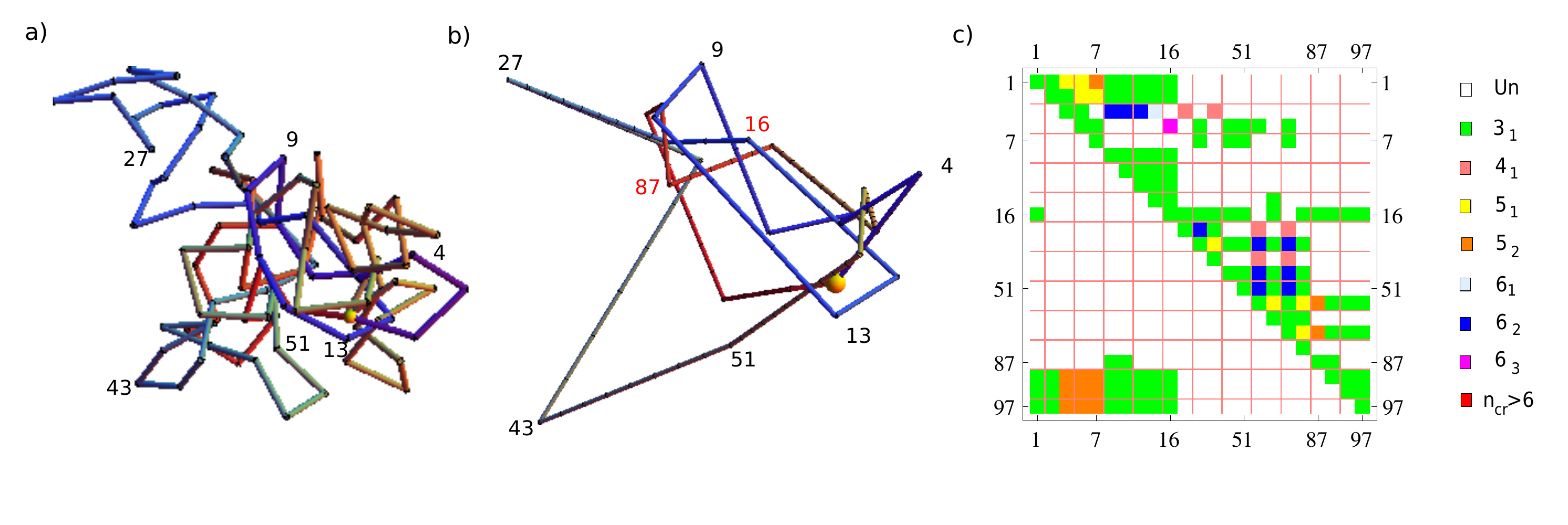}
\end{center}
\caption{ (a) Original and (b) simplified ring after
  the rectification procedure. The original index (numbering) of a
  subset of vertices is shown explicitly. The first vertex is
  highlighted with a yellow sphere. The knot matrix of the simplified
  ring is shown in panel (c). The full knot matrix of the original
  ring is shown in Fig. \ref{fig:max_prob} panel
  (d). }\label{fig:simp}
\end{figure*}

In the attempt to reduce the heavy computational cost of locating 
the knot either in rings or linear chains, several groups have
avoided the extensive topological profiling of all arcs of the ring
and have instead mapped out the topology of a simplified
representation of
it~\cite{Koniaris&Muthukumar:1991b,Taylor:2000:Nature:10972297,Virnau:2005:JACS:127}.

The simplification, or rectification procedure entails the removal of
those ring vertices which can be made collinear with their
neighbouring pair along the ring through a continuous local deformation (morphing) of the ring that does not lead to any edge crossing. Such
rectification operations clearly preserve the topology of the ring and
can considerably reduce the number of ring vertices, and hence the
linear size of the knot matrices.

 Here we discuss the effect of  rectification
  procedure on the two knot localization schemes introduced in
  \ref{knot_localization_schemes}.

In order to ensure the most uniform level of simplification, we
subjected each ring on $N$ edges to several simplification rounds. At each stage of
the procedure we disallow the removal of ring vertices that would
introduce a gap larger than $s$ in the original index of consecutive
surviving beads. Because of the ring periodic
  boundaring conditions, we employ the modulus operation on $s$. By
  starting with $s=2$ we carry out $N$ statistically-independent
  attempts at bead removal (sweep).  Then $s$ is increased by one and
another sweep of vertices elimination is attempted. The procedure is
carried on until no vertex can be further removed within a
sweep. Notice that more beads might be removed by allowing $s$ to
increase even further, but these more aggressive rectifications are
not considered here.

Reducing the number of beads of the ring has two
  distinct effects. First, the linear dimensions of the knot matrix
  are reduced. Second, the topological entanglement of the remaining
  subportions might be different from the one measured for the
  corresponding subportions of the original ring.  Regarding the first
  aspect we recall that we establish the entanglement trapped in the
  arcs of the surviving nodes by closing the arcs with the same ends
  on the original, unsimplified ring. As a consequence the knot matrix
  of the simplified ring is a subset of the original full knot
  matrix, obtained by restricting to the rows and columns pertaining
  to the surviving ring vertices. This procedure is illustrated in Fig
  \ref{fig:simp}.  

Notice that because the simplified knot matrix is a
  subset of the original knot matrix, then the length of
  the shortest knotted arc measured on the simplified ring can not be
  smaller than the length of the shortest knotted arc measured on the
  unsimplified ring. On the other hand no such reasoning can be made
  for the shortest C-knotted arc.

The rectification procedure clearly brings about a
  simplification of the geometrical complexity of the ring. As a
  consequence, the difference between the shortest knotted arc and
  shortest C-knotted arc will likely decrease after rectification but
  it will not be obliterated for sufficiently entangled rings. This is
  illustrated by the rectification of the ring with intermediate
  compactness, whose simplified knot matrix is shown in
  Fig.~\ref{fig:simp}. The shortest knotted arc on the simplified ring
  goes from node 87 to node 16. The inspection of the matrix reveals
  that from the point $(87,16)$, corresponding to the shortest knotted
  arc $\Gamma_{87,16}$, one cannot find a connected path through
  points corresponding to longer and longer trefoil-knotted arcs (even
  disregarding the unknottednedd requirement on the complementary
  arcs) that reach out to the whole ring. This clarifies that
  $\Gamma_{87,16}$ does not correspond to the shortest C-knotted arc,
  and hence the two methods for knot detection do necessarily not
  coincide even after simplification.

\end{document}